# Dampening Long-Period Doppler Shift Oscillations using Deep Machine Learning Techniques in the Solar Network and Internetwork⋆

Rayhaneh Sadeghi[a], Ehsan Tavabi[a,*]

[a]*Physics department, Payame noor university, Tehran, 19395-3697, Iran*

## Abstract

This study aimed to explore the Doppler shift at different wavelengths in the IRIS solar spectrum and analyze the evolution and statistical properties of longitudinal oscillations with damping in chromosphere and transition region bright points (BPs). Understanding the damping mechanism and statistical properties of oscillations in BPs is crucial for gaining insights into the dynamics and energy transport within the solar atmosphere.This study explores the Doppler shift at different wavelengths in the solar spectrum of the Interface Region Imaging Spectrograph (*IRIS*) and implements a comprehensive consideration of Doppler velocity oscillations in the IRIS channels. This comprehensive consideration reveals a propagating periodic perturbation in a large number of chromosphere and transition region (TR) bright points (BPs). To the best of our knowledge, this is the first investigation of the longitudinal oscillations with damping in BPs using comprehensive consideration of the Doppler velocity at various wavelengths. The phenomena of attenuation in the red and blue Doppler shifts of the solar wavelength range were seen several times during the experiments. We utilized deep learning techniques to examine the statistical properties of damping in network and internetwork BPs, as well as active, quiet areas, and coronal hole areas. Our results revealed varying damping rates across different regions, with 80% of network BPs exhibiting damping in quiet areas and 72% in coronal hole areas. In active areas, the figure approached 33%. For internetwork BPs, the values were 65%, 54%, and 63% for quiet areas, coronal hole areas, and active regions, respectively. The damping rate in active regions is twice as high at Internetwork's BPs. The damping components in this study were computed, and the findings show that the damping at all points is underdamped. The observed damping process suggests the propagation and leaking of energetic waves out of TR bright points, potentially contributing to the energy transport from the bright magnetic footpoints to the upper chromosphere, transition region, and corona.

## 1. Introduction

Solar spectrum oscillation studies have been and continue to interest solar physics professionals. The first study on oscillations in coronal bright points (CBPs) by Ugarte-Urra et al. (2004a) discovered a period of 491 seconds in the Solar Ultraviolet Measurements of Emitted Radiation (SUMER) of Solar and Heliospheric Observatory (SOHO) S VI 933.40 $A^{\cdot}$ channel. Intensity shows typical variations in a time scale of 420–650 s that correlated in the transition region Coronal Diagnostic Spectrometer (CDS) of SOHO O VI 629.73 $A^{\cdot}$ and O III 599.60 $A^{\cdot}$ lines, and also the He I 583.34 $A^{\cdot}$ line, while the changes in the coronal line Mg IX 368.07 $A^{\cdot}$

*Corresponding author: Email: etavabi@gmail.com;e_tavabi@pnu.ac.ir;

occurred on a longer time scale, and some oscillatory behavior was detected. They also found damped oscillations peaking at 546 s in the CBPs observed in the O VI 629.73 $A^{\cdot}$ line, but this oscillation did not see in SOHO/CDS He I 584 $A^{\cdot}$ line. Ugarte-Urra et al. (2004b) also reported periods as short as 236 seconds; these kinds corresponded to the chromospheric oscillation range and were associated to the CBP channel footpoints. Tian et al. (2008) used TRACE data taken in the 195 $A^{\cdot}$ and 1216 $A^{\cdot}$ filtergrams and applied Fourier and wavelet transform searching for low-frequency oscillations in CBPs. At CBPs, the oscillatory power is stronger with respect to the surrounding quiet Sun with different periods in different CBPs. The periods Are different in various regions of the same CBP. The periods ranging from 8 to 64 min were reported to last for several cycles in both the coronal part of the CBPs and their chromospheric anchorages. However, the intensity oscillatory behavior relationship with the magnetic reconnection recurrent needs more consideration. The reason for the uncertainty is that the chromospheric periods surpass the 3.7-minute cut-off time for chromospheric sounding, and the acoustic cut-off may grow if the waves move along the inclined magnetic axis. Samanta et al. (2015) illustrated that the recurrent energy release in TR or chromospheric BPs is possibly the cause for the large range of the reported oscillation periods in CBPs. In addition, TR or chromospheric BPs are characterized by a high flux variability in EUV emission. The identified BPs, which are small-scale loops at different temperatures, exhibit significant temporal variability and are not distinguishable between coronal hole and quiet regions Habbal et al. (1990). They are associated with He I dark points and the evolution of photospheric network magnetic fields, with a higher probability of chance encounters during low solar activity Harvey (1985). The association of coronal bright points with the approach and disappearance of opposite polarity magnetic network is dominant, and they are also linked to sites of disappearing magnetic flux Harvey et al. (1986). The time evolution of coronal bright points is strongly correlated with the rise and fall of magnetic flux, and they are associated with small "network flares" Pre's & Phillips (1998). Our objective in this study is to establish the longitudinal oscillations periodicity analysis using different types of Doppler maps already been done on TR or chromospheric BPs brightness oscillations Habbal & Withbroe (1981); Habbal et al. (1990); Ugarte-Urra et al. (2004a); Samanta et al. (2015). We found that this behavior is a common feature in their dynamic evolution. We are aware that the longitudinal Doppler oscillations in the quiet and active Sun and coronal holes, network, and internetwork have been found to be intermittent Sadeghi & Tavabi (2022a) to find the population of in each position; therefore, we are interested in the use of a machine learning technique that allows the study of long time-series with huge databases.

The association of coronal bright points with bipolar magnetic features was established in the first research out on BPs Golub et al. (1974); Golub & Underwood (1977); Sheeley & Golub (1979). In addition, Sadeghi & Tavabi (2022a) found that the Doppler velocity time series is first constructed from the Mg II h & k spectrum at a velocity of 20 km/s, and with the center line of the Mg II h & k peak in this approach; It is necessary to acquire the non-thermal Doppler velocity time slice specifically for the peaks of interest, which in this case are labeled as k3 and h3 for Mg II and the center of spectra for Si IV 1394. By applying Equation (1), the Doppler velocity corresponding to the target speed, approximately 20 km/s, can be determined with a tolerance range of ±10.75 km/s. To construct the Doppler velocity time slice. These individual time slices are then combined to form a comprehensive representation of the Doppler velocities across the selected regions of interest. The Sweet-Parker and Petschek reconnection models are insufficient to explain the typical modest Doppler shift values reported in BPs. A new model is needed to explain these small Doppler shift values in the context of the magnetic reconnection model.

The dampening of these oscillations is one of their most intriguing characteristics. One of the unsolved mysteries in solar physics is the mechanism of heating the outer layers of the sun's atmosphere. Nonetheless, numerous hypotheses and investigations have been made in this field. Landseer reported one of the earliest findings of hydro-magnetic waves damping in 1961, and he linked this attenuation to energy transmission to other layers of the atmosphere Landseer-Jones (1961).

The kink mode waves are one of the most significant and have previously been examined by several researchers Pascoe et al. (2015). These waves exhibit apparent dampening, as Aschwanden (2006); Aschwanden et al. (2003); Aschwanden & Schrijver (2011) have stated several times in their publications. These waves have a damped oscillation in the longitudinal direction. Much research has been done to simulate and model the propagation and damping of transverse waves Pascoe et al. (2010); Pagano & De Moortel (2017). Tiwari et al. (2019) reported the damping of kink waves using a method based on Doppler velocity signal tracking and they discovered that the propagation of kink waves is dependent on the damping frequency.

Slow magneto-acoustic waves are another topic of dampening research. Krishna Prasad et al. (2019) determined that slow magneto-acoustic oscillations are rapidly attenuated in the solar corona. Thermal conductivity, according to De Moortel & Hood (2004), is the cause of this damping. Krishna Prasad et al. (2019) researched the link between damping length and temperature for the first time and determined that the two are not directly related. Throughout their investigation. Krishna Prasad et al. (2014, 2017) also mentioned that damping is frequency dependent. Gupta (2017) believes that temperature-dependent damping may imply that thermal conductivity plays a unique role in the damping of Alfvén waves in the lower corona.

Despite several decades of scientific research dedicated to studying chromosphere BPs, their longitudinal oscillations remained largely unexplored; some classical and brilliant works consider their intensity fluctuations Ugarte-Urra et al. (2004a,b); Tavabi (2018); Sadeghi & Tavabi (2022a). Here is an attempt in that direction, where the focus is on Doppler velocity oscillations on the chromosphere underneath a BP, including network and internetwork all around the Sun observed at spatial and temporal scales that have never been reported before. In particular, this study primarily investigates many of both types of BP Doppler periodical behavior using machine learning and their damping models.

The morphological similarities between the spicules associated with BP indicate the possibility that the network BPs are associated with spicular mass ejections and transient heating of the plasma from chromospheric to coronal temperatures. An immediate investigation of such a connection would, however, require more detailed analysis by combining high-resolution imagery with spectroscopic observations of on-disk BPs with off-limb spicules. Nonetheless, some studies, such as Tavabi (2014), have already illustrated an intriguing connection between spicules in the off-limb chromosphere and brightness in the Ca II H line in Hinode observations.

Statistical studies with coordinated chromospheric and higher-resolution TR spectral observations reveal a damping mechanism with the vast differences between network and internetwork at active regions. In quiet Sun and coronal holes, the damping rate is somewhat higher at network BPs relative to the internetworks. However, this difference is increasing to more than twice in active regions. The complex chromospheric and TR scenery underneath BPs in different areas could show us their critical roles in their formation scenario and their influential contributions to the heating. The goal of this article is to investigate the connection between Doppler velocity damps and the origin of bright points in solar physics. By analyzing *IRIS* Doppler observations, we aim to uncover the damping characteristics and their implications in different solar regions. Additionally, we seek to explore the relationship between damping and energy transmission, shedding light on the mechanisms that contribute to chromosphere and transition region heating. The article aims to provide valuable insights into the role of damping in bright point dynamics and its impact on the solar atmosphere.

The findings of this study have important implications for solar physics and practical applications. By exploring the Doppler shift and longitudinal oscillations in the chromosphere and transition region, we deepen our understanding of solar dynamics and energy transfer processes. The study's results contribute to improved models and simulations, aiding in the interpretation of solar phenomena and enhancing space weather forecasting. Additionally, the analysis of damping properties in different solar regions provides insights into variations across environments, aiding in the classification and characterization of solar features. Overall, this research advances our understanding of the Sun's behavior and its impact on Earth's technological infrastructure.

## 2. Observation

The Interface Region Imaging Spectrograph (IRIS) is a NASA satellite launched in 2013 specifically designed to study the Sun's chromosphere and transition region. This critical interface layer plays a crucial role in channeling energy and plasma into the Sun's outer atmosphere, the corona, and ultimately driving solar wind. IRIS data were used in this study, along with slit jaw images (SJIs) and spectral data from the observatory. *IRIS* spectrum data contains a wide wavelength range. The range of wavelengths from 1332 Å to 2834 Å represents the spectral range covered by the *IRIS*, meaning that the *IRIS* is capable of detecting light or radiation within this range. Within this range, there are various emission lines associated with different elements, which are indicated in the statement. Specifically, the core lines associated with C II, Fe XII, O I, Si IV, and Mg II h k (figure 1). The Mg II h and k lines are highly significant in this study. These lines, observed using the IRIS , offer valuable insights into the temperature, velocities, and behavior of the plasma in these regions. Data with good spectral resolution from the three zones of equatorial coronal holes, the quiet region, and the active region and coronal hole area, were useful for this investigation De Pontieu et al. (2014). From 1332 to 2834 Å, *IRIS* collected spectra in near-ultraviolet (NUV), far-ultraviolet 1 (FUV1), and far-ultraviolet 2 (FUV2). *IRIS* slit-jaw images (SJIs) produced with various filters can provide images focused on the Mg II wing, Mg II k, Si IV 1403 Å, and C II lines (De Pontieu et al. (2014); see the *IRIS* Technical Note 20 for details). Because Mg II h & k has a low FIP, it is typically utilized in plasma at low temperatures and above the minimum temperature $\tau_{500} = 1$. The velocity resolution in *IRIS* spectra is 0.5 km/s.

## 3. Method

In order to improve the organization and clarity of our methodology, we have divided the method section into separate subsections. This allows for a more structured presentation of the different components of our approach. The subsections include "Dataset Preparation", "Deep Learning Model", " Application of the Deep Learning Model on Real Data", and "Postprocessing" . The subdivision of the method section into these subsections provides a more organized and accessible framework to understand and replicate our methodology.

### 3.1. Dataset Preparation

The Doppler shift oscillations over time and above the *IRIS* bright points were explored in this study. To do this, the bright points from SJI 1400 were first separated(Table 1); then their kind was identified in terms of being associated with the network and internetwork using the approach described by Sadeghi & Tavabi (2022a). The equation shown below can be used whenever you need to find out $\Delta v_{Doppler}$:

$$\Delta v_{\text{Doppler}} = -\frac{1}{2}\frac{c}{\lambda_0}\left[(\lambda_{k2v} - \lambda_{k3}) + (\lambda_{k2r} - \lambda_{k3})\right]$$

This equation refers to the speed of light ($c$) and three different wavelengths. The k3 line center wavelength is denoted as $\lambda k3$, while the observed peak wavelength for kv is represented by $\lambda_{k2v}$, and the observed peak wavelength for kr is represented by $\lambda_{k2r}$. In our study, we employed the provided equation to calculate the Doppler velocity ($\Delta v_{\text{Doppler}}$) using specific measurements from IRIS spectra of Mg II. This equation incorporates the speed of light ($c$) and three distinct wavelength values: $\lambda_{k3}$ for the line center wavelength, $\lambda_{k2v}$ for the observed peak wavelength of the kv component, and $\lambda_{k2r}$ for the observed peak wavelength of the kr component. By substituting these values into the equation, we were able to quantify the Doppler velocity associated with the Mg

II spectral lines. This analysis provided valuable insights into the motion and velocity of the observed Mg II lines relative to the observer. Our utilization of the equation with the IRIS spectra of Mg II allowed for a comprehensive investigation of the Doppler velocity and its implications within our research.

In our previous study Sadeghi & Tavabi (2022a), we investigated the relationship between magnetic bright points and underlying magnetic fields. The study utilized data from IRIS and SDO to investigate the Mg II k-line intensity and Doppler velocity above these bright points in the chromosphere and transition region. The identification of bright points was achieved by combining Si IV 1403-Å slit-jaw images with magnetogram information from the HMI. Subsequently, a time-slice analysis and wavelet inspection on the Mg II k resonance lines were conducted to detect oscillation periods, further enhancing our understanding of the characteristics of chromospheric oscillations in magnetic bright points. As a crucial part of our analysis method, we conducted a systematic identification of bright points based on their associated oscillation periods. Specifically, we observed that bright points exhibiting an oscillation period of approximately 180 seconds were typically located in the internetwork regions. In contrast, bright points with an oscillation period of approximately 300 seconds were predominantly found in the network regions. This distinction in oscillation periods and spatial distribution provided valuable insights into the dynamic behavior and localization of these bright points within the solar network and internetwork regions (like panel a and b of figures 2, 3, 4) Sadeghi & Tavabi (2022a). The Doppler velocity time series is first constructed from the Mg II h and Mg II k spectrum at a velocity of 20 km/s and with the center line of the $h_3$ and $k_3$ peak, respectively, in this approach. The time series is then analyzed using wavelets, and the period of Doppler velocity oscillations is estimated based on the results. To ensure the robust identification of bright points in our analysis, we adopted a multi-faceted approach. This approach relied on a combination of observed characteristics and established criteria. The structures classified as bright points exhibited several key characteristics:Enhanced Intensity, Spectral Signatures,Temporal Stability. The classification of bright points relied on established criteria such as, Minimum Intensity Thresholds, Spectral Line Ratios, Spatial Context, Observed Lifetime. This approach aligns with established methods for identifying bright points in solar observations, as documented in previous studies Schmelz et al. (2013); Choudhary & Shimizu (2010); Adami et al. (2000); Zeighami et al. (2020)By incorporating these details, we aim to provide a clear and well-supported justification for the classification of the observed structures as bright points.

Wavelet analysis is employed in this study to extract information on both time and frequency domains while taking into account the uncertainty principle. By convolving wavelets with signals, oscillation power and temporal frequency characteristics can be revealed. The specific wavelet utilized in this research is the Morlet wavelet. It consists of a narrow sine wave modulated by a Gaussian function and can be parameterized by a prime number representing the number of wavelet cycles beneath the Gaussian envelope. The Morlet wavelet exhibits minimal ripple effects due to its smooth edges, enabling more accurate detection of fluctuations and oscillations compared to other wavelet options. Additionally, the Morlet wavelet preserves the frequency of the analyzed oscillations and wave-like disturbances without distortion. In this study, a Morlet wavelet with five sine oscillation periods enclosed within the Gaussian envelope, known as a Morlet 5, is selected, allowing for a reasonable alignment between temporal and frequency analyses with a high level of accuracy. To perform the wavelet analyses, the Torrence & Compo (1998) code was utilized.

This code incorporates considerations for both white and red noise, as described in the technical paper authored by Torrence & Compo (1998). By incorporating these noise components into the wavelet analysis, the researchers ensure robustness and reliability in their findings. These details provide a comprehensive understanding of the wavelet analysis methodology employed in this paper, emphasizing the usage of the Morlet wavelet and the implementation of the Torrence & Compo code. The bright points are classified into six types: network bright points in the active region, internetwork bright points in the active region, network bright points in the quiet area, internetwork bright points in the coronal hole area, and internetwork bright points in the coronal hole area.

When BPs are stated in AR, it means the areas surrounding the sunspot are more active in terms of population, community, and other factors. Bright points that can make a difference in appearance with the quiet sun; for this purpose, we want them to be somewhat close to the penumbra. The timing diagram of Doppler shift oscillations in two spectral lines of Mg II k and Si IV 1394 $\overset{.}{A}$ was created for all bright points (like panel d of figures 2, 3, 4). In the study, the blue and red shifts of the Mg II h k and Si IV 1394 Å spectral lines were computed. Six-time series data were collected for each bright point, and the distances of the blue and red wings from the center of the spectra were calculated. These values, initially based on angstroms, can be alternatively expressed in km/s to represent motion. The reference wavelength was obtained from the center of the spectral lines. Six-time series data were collected for each bright point, linked to the blue and red Doppler shifts of the Mg II h & k and Si IV 1394 $\overset{.}{A}$ spectra (like panel c of figures 2, 3, 4).

The fundamental purpose of this research is to analyze the dampening of Long-Period Doppler Shift Oscillations in the Solar Network and Internetwork in three unique regions: the quiet Sun, the coronal hole, and the active region. Deep learning can help achieve this aim by collecting complex patterns and connections in data that are necessary for understanding the underlying physical processes.

### 3.2. Deep Learning Mode

We have made significant advancements in the identification of bright points in network and internetwork regions by incorporating a deep learning approach. A separate paper provides a comprehensive analysis of the deep learning process and its effectiveness in detecting bright points in these solar regions (Tavabi & Sadeghi (2024)). To train our deep learning model, we employed supervised learning. In this scenario, by altering its internal parameters during training, the model learns to transfer the input time-series data (Doppler shift oscillations) to the right output category (damp, single peak, ascending, or none) Felipe & Ramos (2019); Abdughani et al. (2019). Two *IRIS* data sets, one from the equatorial coronal hole and the other from the Sun's active zone were utilized as training data for the deep learning model. These datasets are composed of a significant amount of data, with each dataset containing more than 500 spectra and covering a duration of over 60 minutes. Four network bright points and four internetwork bright points were chosen from each location. The data of these 16 points were drawn in time intervals, and it was decided if they were damp, a single peak, ascending, or none. To clarify the meaning of these terms, "damp" refers to a reduction in the amplitude of an oscillation over time, and "single peak" refers to an oscillation with a single maximum amplitude (To clarify the meaning of the term "single peak" within the context of oscillation, it specifically refers to a localized portion of the oscillatory pattern where a dominant peak or maximum amplitude occurs. While oscillation typically implies a repetitive or periodic behavior, the term "single peak" is used to describe a distinct and prominent peak within the overall oscillation, highlighting a specific region of heightened amplitude.). "Ascending" refers to an oscillation with rising amplitude, while "none" indicates that the data does not fall into any of the other categories. The training dataset comprised 16 data series. Each data series consisted of two cells: one cell defining the status and another cell containing the main data. Within each cell series, there were four matrices representing the blue and red shifts of Mg II k & h and Si IV 1394 Å. These matrices captured the relevant data for analysis. The deep learning model utilized in this study consisted of four components: Blue Si, Red Si, Blue Mg, and Red Mg. Notably, the data from these 16 series underwent a crucial preprocessing step. They were divided into time intervals and categorized based on their visual characteristics, such as damp, single peak, ascending, or none. This categorization step played a vital role in enabling a more detailed analysis and comprehension of the data. The training dataset consisted of 320 samples, derived from the 16 series. This adequate number of samples contributed to the robustness of the model during the training process. This study's deep learning model has four components: Blue Si, Red Si, Blue Mg, and Red Mg. This model contains 500 hidden layers and categorizes the data as damp, a single peak, ascending, or none. This model has 200 steps and is of the Adam type (see 6 and table 2). The Adam type refers to the optimization algorithm used in the deep learning model, specifically the Adam optimizer. Adam combines the concepts of adaptive learning rates and momentum, allowing the model to dynamically adjust the learning rate for each parameter during training to achieve faster convergence and better performance. A Long Short-Term Memory (LSTM) layer is used in the model, which is a sort of recurrent neural network (RNN) specially intended to handle time-series data and learn long-term relationships. Memory cells, input gates, output gates, and forget gates make up LSTMs. The memory cell stores data throughout time, while the gates regulate the flow of data into and out of the cell. LSTMs are especially important to this work since they excel at processing time-series data, such as the Doppler shift oscillations under consideration. LSTMs are a strong choice among deep learning models since they can capture complicated patterns and relationships in data. Different elements of the data, such as the blue and red shifts of Mg II k & h and Si IV 1394, might be represented by the LSTM components. The gates might govern how the model learns and updates its understanding of the correlations between these features and the damp, single peak, ascending, or none categories. In this study, the supervised learning approach is utilized to train the deep learning model using labeled data. The data were labeled based on the visual characteristics of the time series. Based on the structure of the Doppler shift oscillations over time, we manually reviewed each time series and classified it as a damp, single peak, ascending, or none. This categorization served as the basis for training and assessing the machine learning model. By modifying its internal parameters during training, the model learns to translate input data (time-series Doppler shift oscillations) to the right output category. The model's performance on the test data was evaluated to provide an unbiased assessment of its correctness. The model employed supervised learning to detect patterns in the data. In a total of 200 tests, the accuracy of the model ranged from 58% to 98%, with an average accuracy exceeding 80%. The accuracy measure was used to evaluate how successfully the deep learning model classified data into categories such as damp, single peak, ascending, or none. To determine accuracy, the model's predictions were compared to the actual known classifications of the data. This variability highlights the complexity and inherent challenges in accurately predicting the damping Doppler shift oscillations.

The baseline model served as a reference point for comparison. The baseline model represented the most common category for all data points. The tests involved assessing the model's performance on a new collection of data referred to as the test data. This test data was similar to, but not the same as, the training data used to train the model and the validation data used to fine-tune the model's parameters. Evaluating the model on the test data allowed for an impartial assessment of its performance.

In this study, over 2000 bright points from 20 IRIS data series were input into the deep learning network, and the results were obtained as either dimming or non-dimming. The model's validity was determined by measuring two points in the active region and two points in the coronal hole region. The deep learning model was specifically designed to detect all instances of damping in the Doppler shift in the time series. The damping data could then be fitted using a typical damping model. Accuracy is a measure of how successfully the deep learning model classifies data as a damp, single peak, ascending, or none. To determine accuracy, the model's predictions are compared to the data's actual, known classifications. The baseline, on the other hand, serves as a point of reference for comparison. A baseline model is the most common category for all data points. Class imbalance can pose challenges in classification tasks, particularly when the proportion of samples in different classes is significantly imbalanced. To mitigate the impact of class imbalance on model training, we took several steps to ensure a balanced representation of classes in the training dataset.

First, we employed techniques such as stratified sampling or oversampling/undersampling methods. These techniques adjust the distribution of samples in the training dataset to prevent the model from being biased towards the majority class. By including a balanced representation of both classes, we aimed to enable the model to learn from and differentiate between the classes effectively.

In the testing dataset, we maintained the class proportions consistent with the analysis data distribution to ensure a realistic evaluation scenario. We acknowledge that relying solely on accuracy as an evaluation metric can be misleading in the presence of class imbalance. Therefore, we included additional metrics that provide a more comprehensive assessment of the model's performance.

One such metric is the area under the receiver operating characteristic curve (AUC-ROC). The AUC-ROC metric measures the model's ability to distinguish between the positive and negative classes, regardless of the classification threshold. It provides a robust evaluation even when the class proportions are imbalanced. A higher AUC-ROC score(near 90%) indicates better discrimination between the classes.

Furthermore, we considered precision, recall, and F1-score as evaluation metrics. Precision measures the proportion of correct positive predictions out of all positive predictions, while recall measures the proportion of correct positive predictions out of all actual positive instances. The F1-score is the harmonic mean of precision and recall, providing a balanced measure that considers both metrics.

By utilizing these evaluation metrics, we aimed to capture the performance of the model in terms of correctly identifying both classes, taking into account the potential imbalance between them. This approach allows for a comprehensive and reliable assessment of the deep learning model's performance.

### 3.3. Application of the Deep Learning Model on Real Data

These tests are iterations or trials in which the model is tested on a new collection of data termed the test data. The test data is similar to (but not the same as) the training data, which is used to train the model, and the validation data, which is used to fine-tune the model's parameters. We can gain an impartial assessment of the model's performance by assessing it on test data. More than 2000 bright points from 20 *IRIS* data series were input into the deep learning network, and the results were retrieved as dimming or not. The validity of this model was determined by measuring two points in the active region and two points in the coronal hole region.

### 3.4. Postprocessing

The deep learning model separates all damps in the Doppler shift in time series. The damping data should then be fitted using a typical damping model. Equation 1 gives the general formula for damping, where $y_t$ is the instantaneous amplitude at the time t, A is the initial amplitude of the envelope, $\lambda$ is the decay rate, $\phi$ is the phase angle at t = 0, and $\omega$ is the angular frequency Ruzicka & Derby (1971); Alciatore et al. (2007); Golnaraghi & Kuo (2017). Damping characteristics are determined for each damping diagram using this fitting.

$$y_t = Ae^{-\lambda t}cos(\omega t - \Phi) \tag{1}$$

In our study, we analyzed the damping characteristics by fitting a damping formula to individual damping diagrams. Each damping diagram represents the Doppler velocity over time. The fitting procedure involved determining the best-fit values of the parameters (A, $\lambda$, $\omega$, $\phi$) for each damping diagram. We achieved this by minimizing the discrepancy between the observed data points and the curve generated by the damping formula. This was accomplished using optimization techniques of least-squares fitting. By obtaining the best-fit parameters for each damping diagram, we quantitatively characterized the damping behavior of the phenomenon.

These parameters, including the damping rate ($\lambda$) and angular frequency ($\omega$), provided valuable insights into the underlying dynamics and properties of the damping process.

*3.5. method summary*

In summary, the process of creating a deep learning model involves pre-processing, training, and testing; after determining the research area and organizing the detected bright points, data analysis is done in this study utilizing the deep learning model (Figure 5, 6). The statement describes the uses of deep machine learning to analyze time-series data from *IRIS*).The focus is on analyzing longitudinal waves, which are inferred from the variations in line-of-sight Doppler velocities observed in the Doppler maps. While the primary emphasis is on these longitudinal waves, it is acknowledged that other oscillations, such as transverse waves, may also be present. However, due to the limitations of the data obtained from the IRIS, the apparent motion associated with transverse waves may not be clearly discernible at the determined amplitudes. With regard to intensity observations, while the study primarily focuses on Doppler velocity variations, there is a possibility of accompanying oscillations in intensity, which can be explored through techniques such as wavelet analysis. It identifies the average period of Doppler velocity oscillations for different types of bright points and develops a deep learning model with four components to classify the data, achieving an average accuracy of over 80%. The study provides new insights into the properties and behavior of plasma in the Sun's atmosphere, and the use of deep machine learning in solar physics has the potential to unlock new insights into the behavior and evolution of our closest star.

## 4. Results

In this section, our main focus is on the extraction of Doppler oscillation properties from BP data and the analysis of their statistical parameters. For a detailed visualization and understanding of the Doppler oscillations, I would like to direct the reader's attention to Figures **??** in the previous section. These figures provide clear illustrations of the Doppler oscillation patterns observed in our study. By analyzing these patterns, we aim to derive valuable insights into the statistical properties and characteristics of the oscillations Bright points can be categorized into two groups: non-oscillating and oscillatory. Overall, our findings indicate that the population dynamics of oscillating and non-oscillating solar bright points are complicated. These dynamics are influenced by elements such as magnetic field strength, solar activity, and temporal variability. Both oscillatory and non-oscillatory types of BPs have been observed. Solar bright points that oscillate periodically at particular frequencies are known as oscillating bright points. This oscillation has been reported to occur at various frequencies, ranging from a few seconds to several minutes. On the other hand, non-oscillating solar bright points do not exhibit any observable periodic oscillations. Instead, they are distinguished by their intensity and unchanging appearance.

Research has been done to study the populations of oscillating and non-oscillating solar bright points at networks and internetworks. A study by Tavabi & Sadeghi (2024) found that the population of oscillating bright points was higher in magnetic network regions compared to internetwork regions. This suggests that oscillations occur more frequently in regions with stronger magnetic fields (these results are found in Table 2). The goal of this research is to explore the damping in the time series of Doppler velocity shift in two spectral lines of Mg II h & k and Si IV. We utilized a deep learning model with machine learning methods to isolate damping data from over 2000 samples. We then assessed the data statistically and qualitatively. The data shows that attenuation is visible in the Doppler shift time series of 70% of the points associated with the internetwork, while this figure approaches 60% for the network. Examples of the deep learning model's input data are shown in Figures 2, 3, 4. Table 3 contains more detailed statistical information. Figure 7 presents the interpreted data chosen as the deep learning model's training data. These data are fed into the model as training data. Blue and redshifts of Mg II h & k peaks, as well as Si IV 1394 $A^\cdot$, were utilized as learning data for deep learning. Guidance data, which are single peaks, damps, and ascending, have been found in addition to each series of these data for model training. The status matrix refers to this set of guidance data. Data series were utilized to evaluate the model, and the visual results are shown in Figure 5.

Table 1. Imaging observations

| Imaging Date | Time (UT) | Image Center Coordinates | Raster Step (seconds) | OBSID |
|---|---|---|---|---|
| 2014-01-18 | 13:13:23-14:09:51 | 1",2" | 9.2 | 3820009504 |
| 2014-02-06 | 12:44:17-13:43:49 | 6",32" | 5.1 | 3803257203 |
| 2014-05-03 | 08:30:20-11:29:19 | 22",-79" | 30.9 | 3880112403 |
| 2014-06-28 | 07:58:17-09:59:09 | 23",107" | 5.6 | 3823257653 |
| 2014-09-20 | 23:54:28-02:47:33  +1d | 10",4" | 5.2 | 3820506153 |
| 2014-12-03 | 14:39:17-15:34:01 | -3",7" | 5.5 | 3800257153 |
| 2015-03-13 | 04:59:56-15:28:27 | 9",-154" | 9.2 | 3860109053 |
| 2015-04-08 | 04:57:17-09:33:21 | 45",-119" | 5.3 | 3860107054 |
| 2015-05-05 | 10:01:08-10:59:00 | -0",1" | 5.2 | 3803106853 |
| 2015-08-04 | 17:47:28-19:42:31 | -159",144" | 16.4 | 3620261403 |
| 2015-11-19 | 00:38:20-01:40:06 | 56",168" | 9.4 | 3620258603 |
| 2016-03-30 | 21:29:24-22:31:59 | 56",36" | 2.3 | 3664101603 |
| 2016-05-06 | 12:09:41-15:00:58 | 39",117" | 32 | 3610013603 |
| 2016-06-25 | 15:59:28-18:57:51 | -3",1" | 16.5 | 3600261302 |
| 2016-07-18 | 00:29:10-01:28:52 | -32",24" | 1.4 | 3650203604 |
| 2016-08-03 | 18:09:15-19:59:25 | 145",79" | 5.2 | 3620106803 |
| 2016-09-17 | 00:17:53-01:52:05 | 98",41" | 1.7 | 3644103603 |
| 2016-10-03 | 05:35:19-06:50:26 | -20",50" | 9.4 | 3620259103 |
| 2016-11-02 | 15:39:26-16:46:14 | 85",54" | 1.7 | 3644103603 |
| 2016-12-03 | 10:53:15-11:51:53 | -92",-121" | 1.7 | 3644103603 |
| 2017-02-14 | 07:49:26-09:53:50 | -123",20" | 62.2 | 3690015104 |
| 2017-05-17 | 19:32:57-22:08:28 | -164",156" | 62.2 | 3690015104 |
| 2017-06-06 | 18:58:53-21:58:52 | -131",99" | 16.7 | 3600010103 |
| 2017-08-26 | 19:37:46-22:58:50 | -10",41" | 16.4 | 3620261103 |
| 2017-09-16 | 13:27:50-15:00:25 | 74",35" | 5.6 | 3624087602 |
| 2017-10-01 | 23:44:20-02:18:17  +1d | 77",151" | 61.6 | 3690115104 |
| 2017-12-22 | 13:10:15-14:07:25 | -66",-55" | 3.5 | 3680504103 |
| 2018-01-03 | 11:03:09-11:57:31 | 40",-20" | 5.5 | 3680506103 |
| 2018-05-08 | 20:09:48-23:40:04 | -19",-117" | 3.1 | 3864255602 |
| 2018-06-18 | 16:49:50-21:59:15 | 66",72" | 16.5 | 3600260003 |
| 2018-07-03 | 07:29:37-09:29:04 | -2",5" | 3.3 | 3600504802 |
| 2018-08-03 | 16:54:28-19:48:09 | -1",3" | 5.2 | 3620106803 |
| 2018-09-18 | 04:19:19-06:01:09 | 94",30" | 9.3 | 3620508703 |
| 2019-07-04 | 09:44:33-12:04:08 | 2",-6" | 9.3 | 3620258602 |
| 2019-07-25 | 00:52:32-03:57:26 | -0",1" | 9.3 | 3620258703 |
| 2019-08-15 | 21:30:00-22:49:29 | -1",2" | 9.4 | 3620258704 |
| 2019-12-12 | 17:30:21-18:27:28 | 167",139" | 9.2 | 3620108803 |
| 2020-04-20 | 08:32:36-09:56:15 | -9",2" | 9.6 | 3620008803 |
| 2021-07-04 | 16:59:50-22:57:17 | -1",-1" | 16.5 | 3624010103 |
| 2022-08-16 | 22:49:36-01:00:54  +1d | -135",64" | 9.4 | 3660259102 |
| 2022-10-16 | 04:05:55-05:05:25 | -135",-98" | 5.5 | 3620257204 |
| 2022-11-06 | 17:25:37-17:33:27 | 12",-148" | 2.2 | 3624601703 |
| 2023-01-19 | 10:35:19-11:35:36 | -70",-161" | 9.4 | 3660259103 |
| 2023-02-05 | 22:49:24-23:58:49 | 11",-101" | 9.4 | 3660259102 |

Table 2.

| | Name | Type | Activation | Learnables | Total Learnables | States |
|---|---|---|---|---|---|---|
| 1 | sequenceinput | sequence input | 6 | - | 0 | Hidden State 500x1 Cell State 500x1 |
| 2 | lstm | LSTM | 500 | input weights 2000x6 recurrent weights 2000x500 Bias 2000x1 | 1014000 | - |
| 3 | fc | Fully connected | 6 | weights 6x500 Bias 6x1 | 3006 | - |
| 4 | softmax | Softmax | 6 | - | - | - |
| 5 | classoutput | Classification Output | 4 | - | - | - |

Table 3. statistical detail of deep learning results

| region | point type | damp % | ascend % | single peak % |
|---|---|---|---|---|
| coronal hole | network | 72 | 55 | 18 |
| | internetwork | 54 | 37 | 57 |
| active | network | 33 | 29 | 25 |
| | internetwork | 63 | 62 | 10 |
| quiet | network | 80 | 52 | 21 |
| | internetwork | 65 | 30 | 54 |

Figure 5 shows a time series of the Si IV 1394 $A^\cdot$ spectrum. A sign of attenuation can be observed in this image. In this diagram, the red and blue wings represent the Doppler velocity shift. It can be observed that a blue shift corresponds to redshift attenuation. Damping is evident in the graphs presented.

The guide data and projected data are presented together in this graphic. This information is related to the redshift of the Si IV 1394 $A^\cdot$. The predictions for two red and blue shifts are displayed in Figure 3. Blue represents attenuation, orange represents a single peak, and purple represents an increasing tendency.

The results of a wavelet analysis on Doppler velocity oscillations for network and internetwork bright points are displayed.

The data obtained from the Mg II h & k peaks and Si IV 1394 $A^\cdot$ spectral lines in the Sun's atmosphere were analyzed using deep learning. The blue and red shifts of these spectral lines, as well as guidance data that categorizes the oscillations into single peaks, damps, and ascending, were used as learning data for deep learning. The status matrix refers to this set of guidance data, which is used to train the deep learning model. Figure 7 presents the interpreted data chosen as the deep learning model's training data, which is fed into the model for training. Data series were utilized to evaluate the model, and the visual results are shown in Figure 5. This figure demonstrates a time series of the Si IV 1394 $A^\cdot$ spectrum, which shows a sign of dampness. The guide data and projected data are drawn together in this graphic, and this information is connected to the Si IV 1394 Å redshift. In panel (c) of Figure 5, we present a plot that combines the guide data and projected data. The guide data represents the observed or labeled data, specifically connected to the Si IV 1394 Å redshift. On the other hand, the projected data is generated by a machine learning model and represents the predicted values. By drawing together the guide data and projected data in this graphic, we aim to assess the accuracy of the machine learning model's predictions in capturing the redshift behavior of the Si IV 1394 Å spectrum. This comparison provides valuable insights into the model's performance and its ability to replicate the observed redshift characteristics. Overall, the combined visualization of the guide data and projected data in panel (c) offers a valuable analysis of the relationship between observed redshift characteristics and the model's predictions, shedding light on the accuracy of the machine learning model's performance in this context. The predictions for two red and blue shifts are displayed, where blue represents attenuation, orange represents a single peak, and purple represents an increasing tendency. The statement also mentions that the results of a Doppler velocity oscillations wavelet analysis for network and internetwork bright points are displayed in a separate figure. Overall, by use of deep learning techniques to analyze complex data from the Sun's atmosphere, and provides insights into the behavior and evolution of plasma in this region. By utilizing spectral data and guidance data for model training, we are able to develop a deep-learning model with high accuracy. The visual results of the model evaluation also highlight the presence of dampening and oscillations in the Si IV 1394 $A^°$ spectrum.

$$f = \omega/(2\pi) \tag{2}$$

$$\tau = 1/\lambda \tag{3}$$

$$Half-life = ln(2)/\lambda \tag{4}$$

$$\zeta = \lambda \sqrt{\lambda^2 + \omega^2} \tag{5}$$

$$Q = 1/(2\zeta) \tag{6}$$

In order to gain a deeper understanding of damping properties, it is necessary to define several components. The components include the time constant, damping frequency, half-life, damping ratio, and quality factor (refer to equations 1- 6). The damping ratio, represented by $\zeta$ (zeta), is a parameter that characterizes the damping properties of a system. It measures the relative strength of the damping force in relation to the critical damping value. A higher damping ratio indicates a greater damping effect, whereas a lower damping ratio suggests a weaker damping influence. Furthermore, we consider the quality factor, also referred to as the Q-factor, which is a dimensionless parameter that characterizes the quality or efficiency of a resonant system. It represents the ratio of energy stored in the system to the energy dissipated per cycle of oscillation. The quality factor offers insights into the sharpness and selectivity of resonant systems and is inversely proportional to the damping ratio. Both the damping ratio and quality factor are fundamental in the analysis of dynamic systems across various engineering disciplines. They play a crucial role in studying mechanical vibrations, electrical circuits, and control systems, offering valuable insights into system behavior, stability, and response characteristics Rivière (2003); Iglesias (2000). Given these parameters, a qualitative damping analysis may be performed. Table 2 shows the average of damp components in various regions. In this context, the term repeat frequency refers to the average time interval between successive occurrences of damping. It represents the average period at which the damping phenomenon recurs within the system. Understanding the repeat frequency provides valuable insights into the cyclic nature of damping events and the regularity with which they occur. It allows researchers to gain a deeper understanding of the system's behavior and the temporal patterns associated with damping. The time constant reflects the amount of time that a damped wave decreases its amplitude by the factor of e. This component is a time metric for wave damping. The greater this component, the longer it takes for a resonance to terminate.

Another crucial damping component in this study is the damping ratio. This component is a mathematical representation of the amount of damping in each cycle. The type of damping can be determined using this component. We shall notice the underdamped state if this number is between 0 and 1. The lower the damping of the oscillation, the closer this value is to zero. The Q factor is another measure that may be used to assess damping. A high Q suggests that damping is sluggish in comparison to oscillation. In our research, we calculated all the parameters presented in Table 2. The quality factor, which is inversely proportional to the damping ratio, offers valuable insights into the sharpness and selectivity of resonant systems. A higher quality factor indicates a more selective and less damped system, while a lower quality factor suggests a broader and more damped system.

Table 4. average of damp components

| region | point type | $f$ | $\tau$ | half-life | $\zeta$ | $Q$ | repeat frequency |
|--------|-----------|-----|--------|-----------|---------|-----|------------------|
| **coronal hole** | network | 0.8 m | 3500 | 2400 | 0.34 | 1.47 | 1500 |
| | internetwork | 1.6 m | 2500 | 1700 | 0.13 | 3.84 | 1000 |
| **active** | network | 2.5 m | 2000 | 1400 | 0.33 | 1.48 | 4000 |
| | internetwork | 1.7 m | 4800 | 3300 | 0.44 | 1.14 | 11000 |
| **quiet** | network | 1 m | 3000 | 2000 | 0.32 | 1.56 | 1200 |
| | internetwork | 1.5 m | 1900 | 1300 | 0.17 | 2.94 | 1000 |

By studying the damping ratio and quality factor, we gain insights into the behavior, stability, and response characteristics of dynamic systems.

Table 3 presents the average of the damping components in various regions, allowing for a qualitative analysis of damping. The term "repeat frequency" refers to the average time interval between successive occurrences of damping. It provides insights into the cyclic nature of damping events and the regularity with which they occur. Understanding the repeat frequency helps researchers grasp the temporal patterns associated with damping, providing a deeper understanding of the system's behavior.

The time constant is another important parameter that reflects the rate at which a damped wave decreases its amplitude by the factor of e. It serves as a time metric for wave damping, with a larger time constant indicating a longer duration for the damping process to terminate.

Additionally, the damping ratio is a crucial component in our study. It mathematically represents the amount of damping in each cycle and determines the type of damping present. An underdamped state is observed when the damping ratio is between 0 and 1, with lower values indicating less damping. On the other hand, the Q factor is another measure used to assess damping. A high Q factor suggests that damping is sluggish compared to oscillation.

By considering these parameters, we gain a comprehensive understanding of the damping characteristics within the system, including cyclic patterns, time duration, and the type of damping. These insights contribute to the overall analysis and interpretation of the study's findings.

## 5. discussion

Energy is required to heat the solar corona to temperatures in the millions of degrees Axford et al. (1999) and to accelerate particles of the solar wind to hundreds of kilometers per second Cranmer et al. (2007). Longitudinal waves, which are oscillations of frozen plasma traveling inside the magnetic field, have been proposed as a potential mechanism for transporting magneto-aquatic energy upward along the solar magnetic flux tubes into the transition region (TR) and corona. Tomczyk et al. (2007) observations of amplitudes of just 0.5 km/s, the energy delivered by the corona waves is inadequate to account for the energy necessary to propel the fast solar wind or counteract the corona's radiative losses, which generally range from 100 to 200 $W/m^2$ Hollweg (1973). Our spectral data from the transition area (the layer of rapidly increasing temperature between the chromosphere and the corona) and the corona, on the other hand, indicate oscillations have been observed in the Sun's outer atmosphere. The amplitudes of the outward-propagating Doppler motions documented in our investigation range from 10 to 35 km/s and periods of the order of 180–550 s throughout the network and internetwork BPs (compatible with recent investigations by Sadeghi & Tavabi (2022b)), and should have sufficient energy to accelerate the fast solar wind and heat the quiet corona.

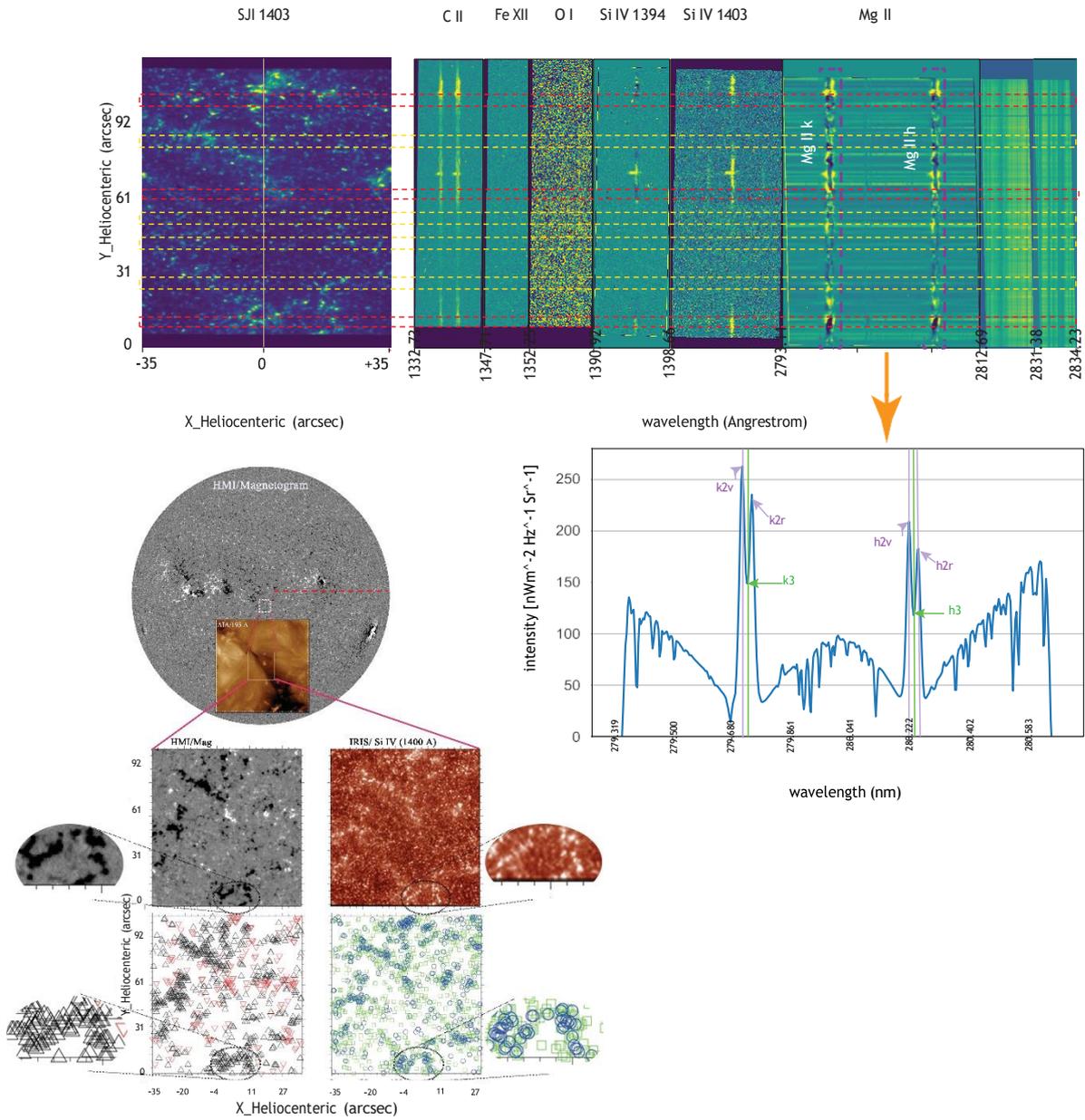

Fig. 1. *IRIS* rasters are drawn from 1332 $\AA$ to 2834 $\AA$ in this illustration. This interval contains core lines associated with C II, Fe XII, O I, Si IV, and Mg II h & k. The exact date and time of the IRIS observations were October 14, 2016, at 20:25. In this panel, the red dashed lines are network bright point areas, and the yellow dashed lines are internetwork bright point areas. At the magnified figure of SDO/HMI related to October 14, 2016, at 20:25, the network and internetwork BPs are illustrated on aligned HMI and SJI images. Below the HMI image, the red triangles represent network BPs and the black triangles represent internetwork BPs. Below the SJI image, the blue circles represent network BPs and the green squares represent internetwork BPs.

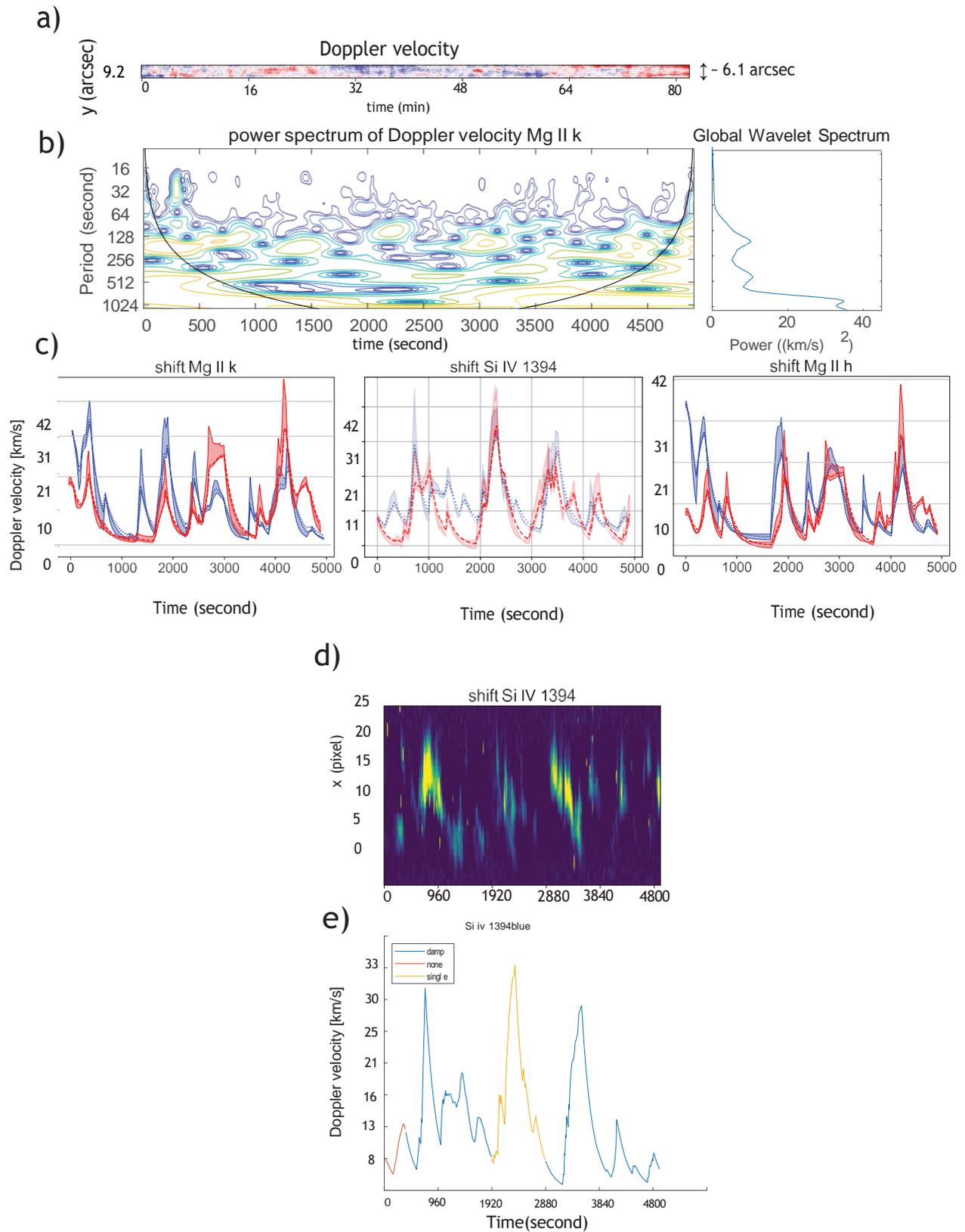

Fig. 2. Internetwork bright points of coronal hole: **a**, Doppler velocity time slice (red is **+20**km/s, blue is -20km/s ,and white is 0 km/s.) **b**, wavelet analysis results for Doppler velocity at 20km/s with center line of Mg II k minima. **c**, red and blue shift of Doppler velocity diagram for Mg II k & h and Si IV 1394 $\AA$. **d**, this figure illustrated the Si IV 1394 $\AA$ Doppler shift vs time; at this time slice, oscillations are obvious. **e**, this diagram is related to the predicted status of Doppler velocity oscillations.

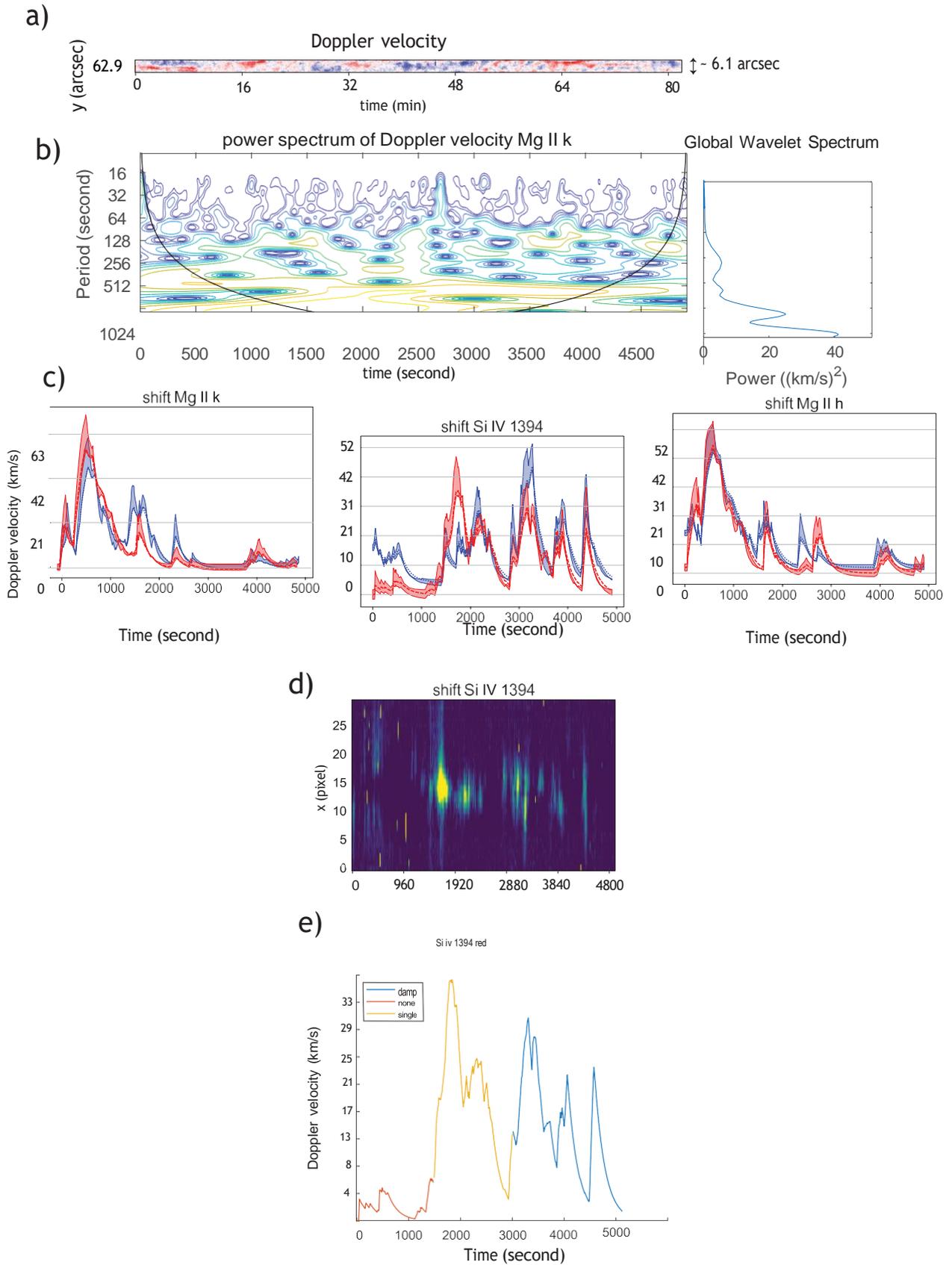

Fig. 3. Same figure 2 for network bright points of coronal hole

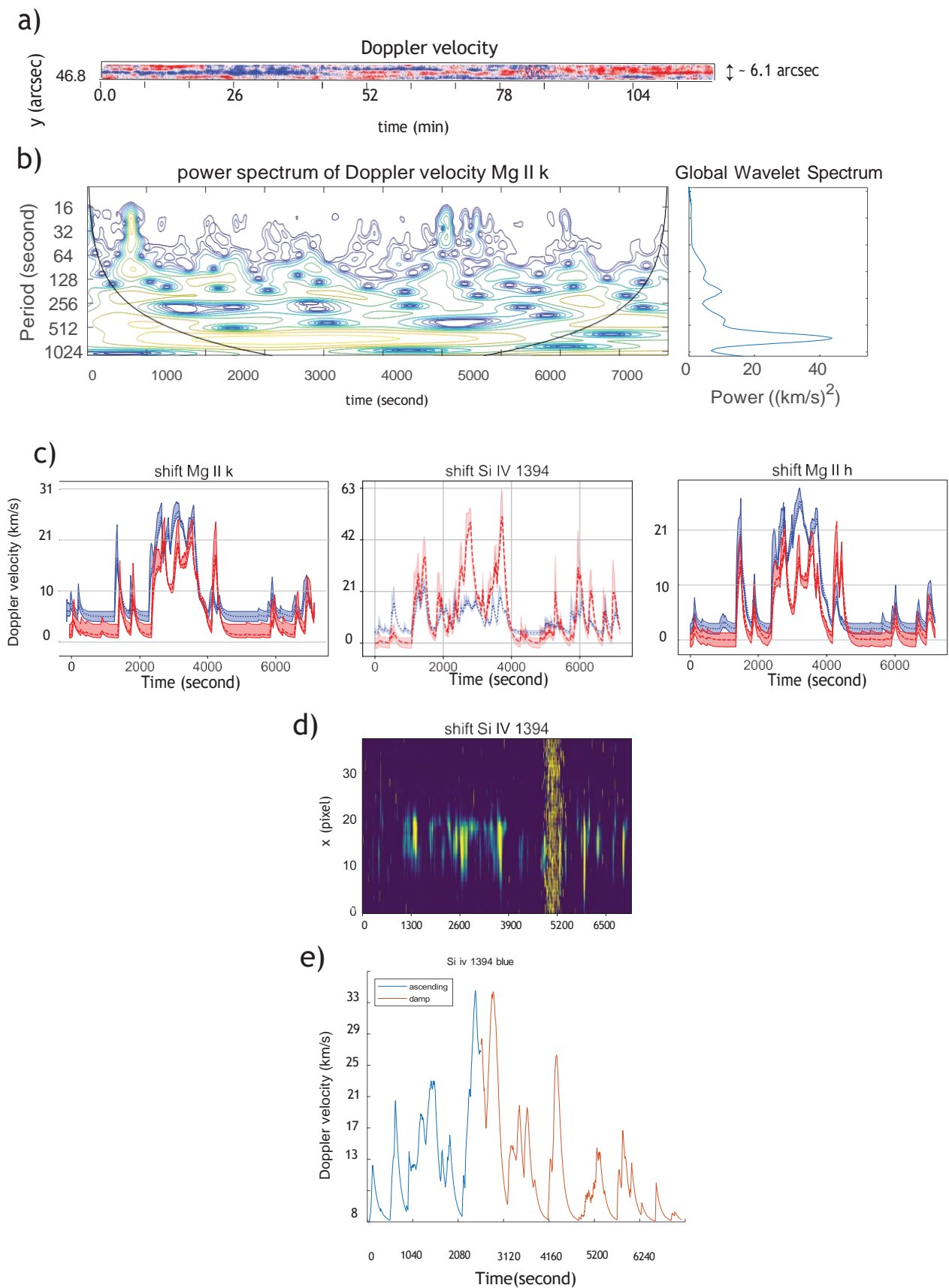

Fig. 4. same figure 2 for internetwork bright points of active regions

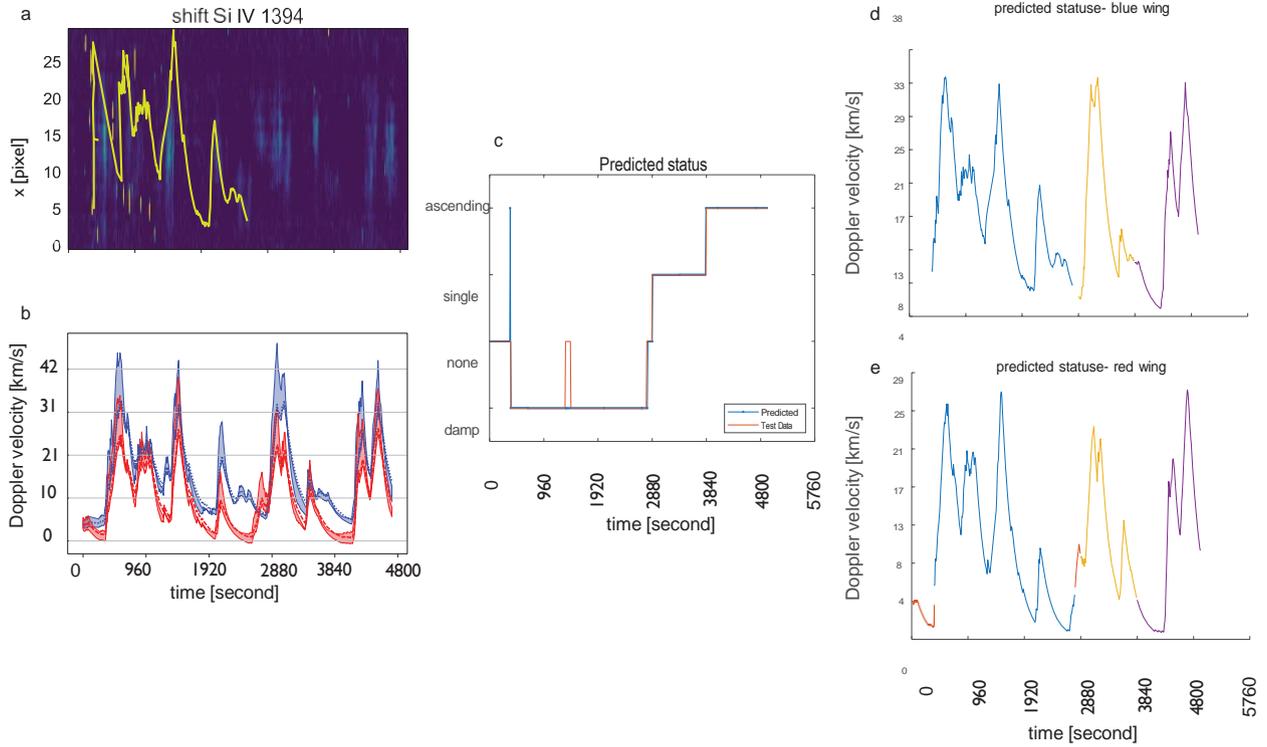

Fig. 5. **a**, The time series of the silicon 1394 $\mathring{A}$ spectrum is depicted in this figure. A sign of dampness may be detected in this image. **b**, The Doppler velocity shift is depicted in this diagram by the red and blue wings. As can be observed, a blue shift corresponds to the redshift dampening. **c**, The guide data and projected data are drawn together in this graphic. This information is connected to the Si IV 1394 $\mathring{A}$ redshift. **d,e**, The predictions for two red and blue shifts are displayed. Blue represents attenuation, orange represents a single peak, and purple represents an increasing tendency.

Solar spectra have been investigated and researched several times throughout the history of solar science Bocchialini et al. (2011); Tavabi (2014); Ahangarzadeh Maralani et al. (2017). Several of these studies have mentioned damping in various solar spectra Duckenfield et al. (2021); Cranmer & Van Ballegooijen (2003); De Moortel & Hood (2003, 2004). Nevertheless, in the analysis of Doppler velocity shifts in terms of time, signal attenuation is seldom emphasized, and generally, only other features of the wave connected to a periodic and undamped wave are examined.

We find as well, that the majority of these networks and internetwork BPs oscillations have a damped nature with different percentages of damping in different regions. We notice that all these damps lie in the period range of the chromospheric and TR oscillations at internetworks, and close to photospheric oscillations at networks. We also come to the conclusion that some of the observed network BPs in the chromospheric network are probably the footpoints or possibly cross sections of the spicules that make up chromosphere BPs (also see Tavabi (2014)). The lower chromosphere below the network BP shows a significant relationship with jet-like structures with a number of peculiar features concerning its density and the spicules in it, however the network BP underneath a spicule, with the aim of unraveling the unexplored impact of BPs on the spicular activity and vice-versa.

We suggest that these oscillations are the same as those seen in the chromospheric network and that a fraction of the network bright points is most likely the cool footpoints of flux tubes comprising coronal bright points. These oscillations are interpreted in terms of global acoustic modes of the closed magnetic structures associated with BPs Madjarska (2019); Tavabi (2014); McAteer et al. (2002).

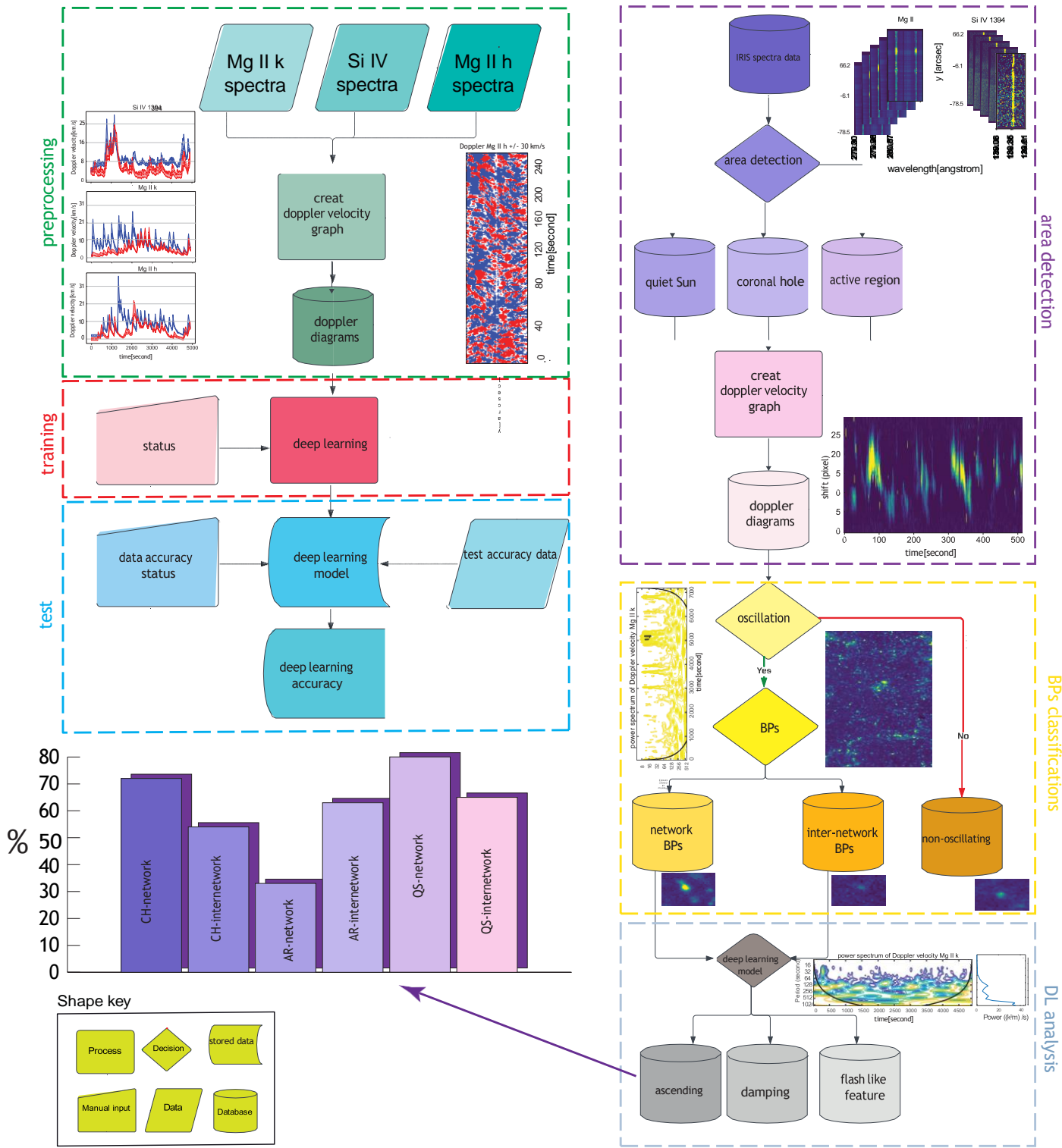

Fig. 6. The deep machine learning examination of the finest time-series *IRIS* rasters validates the rhythmic regime of longitudinal waves discovered using Doppler maps. This research looked at the Doppler shift oscillations over time and above the *IRIS* bright spots. For each bright point, four-time series data were obtained and matched to the blue and red Doppler shifts of the Mg II h & k and Si IV spectra. The average period of Doppler velocity oscillations for network and internetwork points is 300 and 202 seconds, respectively, and bright points are now classified into six types: network bright points in an active region, internetwork bright points in an active region, network bright points in a quiet region, internetworks in a coronal hole area, and internetworks in a coronal hole region. This work employed 16 training data series to build a deep learning model with four components: Blue Si IV, Red Si IV, Blue Mg II, and Red Mg II. The model has 500 hidden layers and classifies data as moist, a single peak, rising, or none. The model's accuracy ranged from 58% to 98%, with an average of more than 80%. The model's validity was established by measuring two points in the active zone and two points in the coronal hole region.

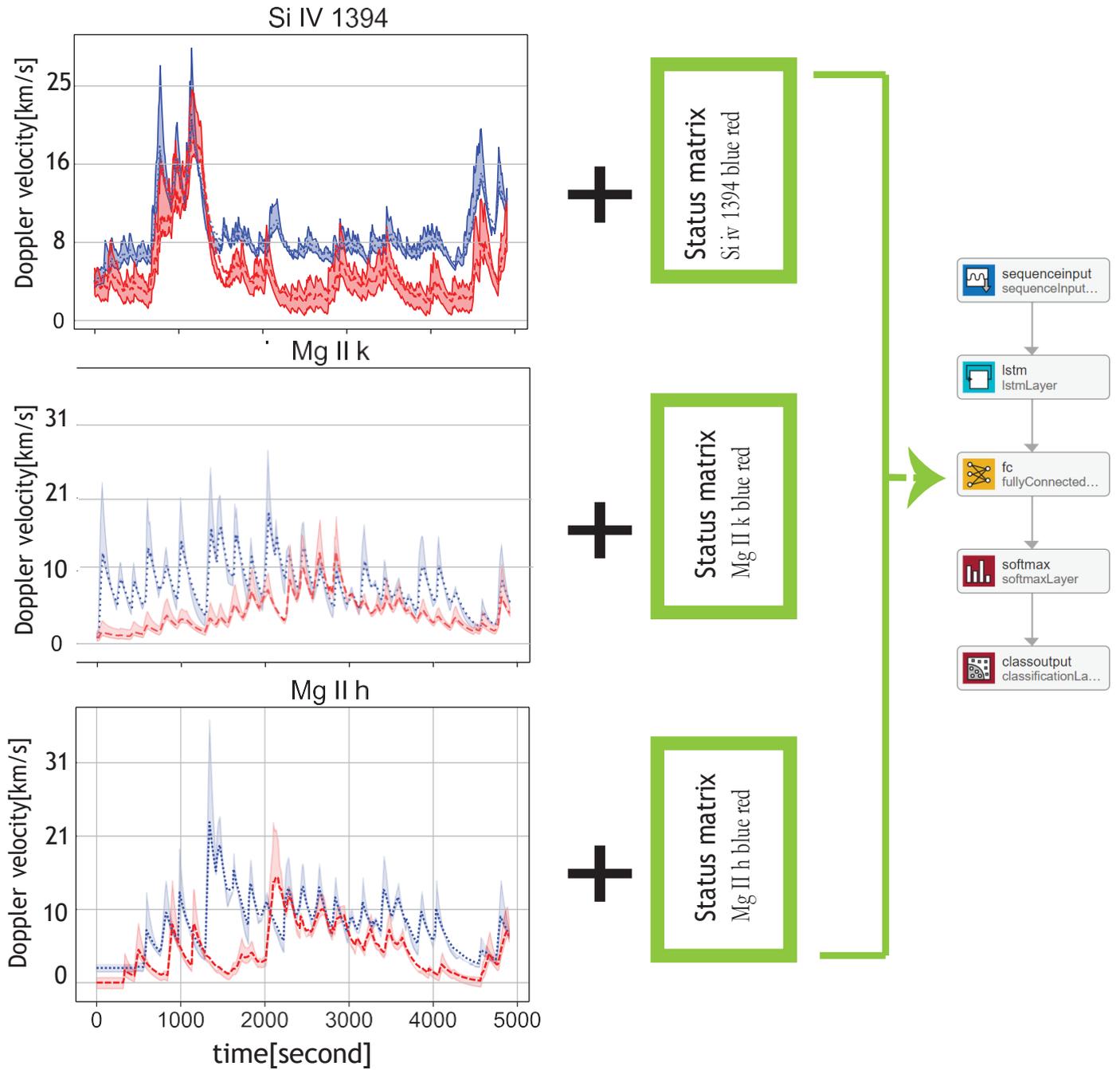

Fig. 7. **a**, Blue and red shifts of Mg II h & k peaks, as well as Si IV 1394 $\AA$, were utilized as learning data for deep learning. Guidance data, which are single peaks, damps, and ascending, have been found in addition to each series of these data for model training. The status matrix refers to this set of guidance data and categorizes the oscillation into single peaks, damps, and ascending.

Wave damping may be seen in numerous types of bright points in various areas, according to qualitative investigations done on various locations and types of bright points, with the exception that these damping appeared to be distinct in terms of statistics and damping characteristics. As a result, we recommended that more data should be examined for future research. So we moved to deep learning to evaluate massive amounts of data.

Therefore, the oscillatory Doppler shifts are interpreted in terms of (standing or propagating) waves in BPs by an impulsive injection of energy at the event's start. First, the site where the oscillation component occurred in the NUV spectrum was brighter after the first Doppler shift, suggesting that this BP was related to the *IRIS* events. The term *IRIS* events refer to a variety of solar events or phenomena that have been observed and studied by *IRIS*. These events can encompass a range of phenomena, including but not limited to solar flares, microflares, coronal mass ejections (CMEs), prominences, filament eruptions, and other dynamic processes occurring in the interface region. Second, the *IRIS* Doppler maps show periodical features in time-slice diagrams, such as periods, damping times, the presence of two components in antiphase, and typical significant initial shifts. Also, they occur at the same spatial positions. All this strongly suggests that their topology and morphology were the same during the events. The statistical analysis according to the deep machine learning output reveals apparent differences between damping rates of *IRIS* BPs in active regions and the quiet Sun (and coronal holes), which are outlined in table 3. Underneath the network BPs, the field strength is distinctly more muscular compared to the internetworks BPs Sadeghi & Tavabi (2022a), where the rapid expansion of the weaker field lines likely leads to different TR impacts.

The damping mechanism of deep machine learning discovered BPs longitudinal oscillations provide clues to the mechanism of Chromosphere and TR heating. We determine the scaling of the damping time with the parameters of the BPs observed in NUV by the *IRIS* sit and stare raster data. Using the observed parameters of the damped oscillating BPs, we determined the dependence of the damping time with the BPs oscillation period, and the Doppler velocities (see Tavabi & Sadeghi (2024)).

Due to the limitations of the highest resolution and diffraction limit of *IRIS* observations, the grouping and cluster behavior of multiple mini-BPs in the network or internetwork areas are not considered in all IRIS rasters and simulated data. The highest possible *IRIS* spatial resolution is approximately 0.33 arcsec; it is clear that some BPs may have smaller diameters.

According to the statistical results of Table 3, damping is visible in 80% of the bright points of the network and 65% of the bright points of the internetwork in the quit Sun region; these values are 33%, 63% for the active region and 72%, 54% for the coronal hole, respectively. The Doppler maps show waves were observed on BP features at the solar disc that protrudes into the higher hot layer.

According to Table 4, in the areas of Coronal Hole and Quiet Sun, the number of damps of the network's bright points is more than the number of network bright points, while in the active area, the ratio is the reverse.

When studying the damping components in the Doppler velocity shift, as shown in Figure 5, the damping length strongly depends on the specific type of points and their area. This issue tends to rely on the origin and root of these bright points. Sadeghi & Tavabi (2022a) explored the origin of the *IRIS* bright points. According to their paper, internetwork bright points have photospheric origins, while network bright points have chromospheric origins.

As a result, it appears that the Doppler velocity damps are connected to the origin of the bright points. According to the damping ratio, the active region appears to have the most damping among the tested locations. Additionally, in the quiet Sun and coronal hole areas, the bright points of the network are damper, but in the active area, the bright points of the internetwork have a higher rate of decay.

The faster an oscillation is damped (a smaller time constant), the faster it transmits energy to the environment, which can create shock waves in higher layers.

The energy of the bright points of *IRIS* was investigated. It was concluded that most of the energy of the network's bright points is transmitted to higher levels, while most of the energy of the internetwork's bright points is returned to lower layers. This issue may be directly related to the effects of damping at the network and internetwork bright points.

In summary, *IRIS* Doppler maps display periodic patterns in time-slice diagrams such as frequencies, damping times, the existence of two components, and typical significant beginning shifts. The BPs longitudinal oscillation clues provide the mechanism of chromosphere and TR heating. The damping time is scaled using the characteristics of the BPs seen in NUV by the *IRIS* sit and stare raster data; the number of damps of the network's bright points is more than the number of network bright points in the coronal hole and quiet Sun regions, whereas the ratio of damping oscillation is reversed in the active area. The damping length is greatly dependent on the exact kind of points and their area.

## Acknowledgements

We sincerely thank the referees for their valuable feedback and insightful comments, which greatly improved this article. We extend our heartfelt appreciation to the National Aeronautics and Space Administration (NASA) for granting us access to the data and resources from the Interface Region Imaging Spectrograph (IRIS) mission, the Solar Dynamics Observatory (SDO), and the Helioseismic and Magnetic Imager (HMI).

## Data Availability

The data utilized in this article is available at the Interface Region Imaging Spectrograph (IRIS) mission website (https://www.iris.lmsal.com). The specific details of the data used, including the observation dates, instrument settings, and data processing procedures, are mentioned in the text of the article.

Researchers and interested parties can access the data by visiting the IRIS website and following the provided instructions for data retrieval. The IRIS mission provides a comprehensive data archive, ensuring the availability of the data used in this study to the scientific community.

Please refer to the article text for further information on the specific data sets used and their corresponding details.

## Appendix: Deep Learning Model Architecture

The deep learning model utilized in this research is composed of four components: Blue Si, Red Si, Blue Mg, and Red Mg. Each component handles different aspects of the data, representing the blue and red shifts of Mg II h & k and Si IV 1394 Å. The model incorporates a Long Short-Term Memory (LSTM) layer, which is a type of recurrent network (RNN) designed to process time-series data and capture long-term relationships. The architecture of the deep learning model is as follows:

1. Input Layer:

- The input layer receives the time-series data of the Doppler shift oscillations.

- The input data is divided into four components: Blue Si, Red Si, Blue Mg, and Red Mg.

2. LSTM Layer:

- The LSTM layer is responsible for processing the time-series data and capturing long-term dependencies.

- It consists of memory cells, input gates, output gates, and forget gates.

- The memory cells store information across time, while the gates control the flow of data into and out of the cells.

- The LSTM layer enables the model to learn complex patterns and relationships in the data.

3. Hidden Layers:

- The deep learning model contains 500 hidden layers.

- These layers facilitate the extraction of higher-level features and representations from the input data.

4. Output Layer:

- The output layer categorizes the data into four classes: damp, single peak, ascending, or none.

- The model predicts the category that best represents the input time-series data.

5. Adam Optimizer:

- The deep learning model utilizes the Adam optimizer for training.

- The Adam optimizer is an optimization algorithm that dynamically adjusts the learning rate for each parameter during training.

- It combines the concepts of adaptive learning rates and momentum, leading to faster convergence and improved performance.

*Training Process*

The deep learning model is trained using supervised learning. The training data consists of two *IRIS* datasets, one from the equatorial coronal hole and the other from the Sun's active zone. Each dataset contains over 500 spectra and covers a duration of more than 60 minutes. From each location, four network bright points and four internetwork bright points are chosen, resulting in a total of 16 data points. The data from these points are divided into time intervals and categorized as damp, single peak, ascending, or none based on their visual characteristics. During training, the internal parameters of the deep learning model are adjusted iteratively to minimize the difference between the predicted categories and the actual categories of the training data. The model learns to associate the input time-series data with the correct output category using gradient descent optimization.

*Performance Evaluation*

The performance of the deep learning model is evaluated using test data. The accuracy of the model is calculated by comparing its predictions with the known classifications of the test data. The accuracy measures the model's ability to correctly classify the data into the damp, single peak, ascending, or none categories. The model achieves an average accuracy of over 80% in 200 tests, with accuracies ranging from 58% to 98%. To evaluate the performance of the deep learning model, a separate test dataset was prepared, consisting of samples that were not used during the training or validation stages. The test dataset was carefully curated to encompass a diverse range of scenarios and to provide a realistic assessment of the model's performance in real-world conditions.

The testing procedure involved applying the trained deep learning model to the test dataset and comparing the model's predictions against the ground truth labels. This allowed for an objective evaluation of the model's ability to accurately classify the samples according to the desired criteria.

It is important to note that the accuracy range reported in our study, varying from 58% to 98%, reflects the inherent complexity of the classification task and the diverse nature of the test dataset. This wide range highlights the challenges associated with accurately classifying certain instances and emphasizes the need for further investigation and analysis to better understand the factors contributing to the performance variations.

*Model Validity*

To assess the validity of the deep learning model, more than 2000 bright points from 20 *IRIS* data series are input into the model. The model predicts whether the bright points correspond to dimming or not. Additionally, two points in the active region and two points in the coronal hole region are measured to evaluate the model's validity in different solar regions. This deep learning model, with its LSTM layer and multiple hidden layers, effectively captures the patterns and connections in the Doppler shift oscillations of the solar network and internetwork. It provides a valuable tool for analyzing the dampening of Long-Period Doppler shift oscillations in different solar regions.